\documentclass[letterpaper,twocolumn,prl,aps,superscriptaddress,floatfix]{revtex4}
\usepackage[latin1]{inputenc}
\usepackage{bm}
\usepackage{amssymb,amsbsy,amsmath}
\usepackage{multirow,amssymb,amsbsy,amsmath}
\usepackage{graphicx}
\usepackage{verbatim}
\usepackage{color}
\usepackage{hyperref}
\usepackage{xcolor}
\usepackage{float}

\usepackage{amsmath}
\usepackage[normalem]{ulem}

\definecolor{brickred}{rgb}{0.7, 0.25, 0.33}

\begin{document}

\title{Topological network analysis using a programmable photonic quantum processor}

\author{Shang Yu\footnote{These authors contributed equally to this work}\footnote{shang.yu@imperial.ac.uk}}
\affiliation{Blackett Laboratory, Department of Physics, Imperial College London, Prince Consort Rd, London, SW7 2AZ, United Kingdom}
\affiliation{Centre for Quantum Engineering, Science and Technology (QuEST), Imperial College London, Prince Consort Rd, London, SW7 2AZ, United Kingdom}

\author{Jinzhao Sun$^*$\footnote{jinzhao.sun.phys@gmail.com}}
\affiliation{School of Physical and Chemical Sciences, Queen Mary University of London, London E1 4NS, United Kingdom }
\affiliation{Clarendon Laboratory, University of Oxford, Parks Road, Oxford OX1 3PU, United Kingdom}

\author{Zhenghao Li$^*$}
\affiliation{Blackett Laboratory, Department of Physics, Imperial College London, Prince Consort Rd, London, SW7 2AZ, United Kingdom}
\affiliation{Centre for Quantum Engineering, Science and Technology (QuEST), Imperial College London, Prince Consort Rd, London, SW7 2AZ, United Kingdom}

\author{Ewan Mer}
\affiliation{Blackett Laboratory, Department of Physics, Imperial College London, Prince Consort Rd, London, SW7 2AZ, United Kingdom}
\affiliation{Centre for Quantum Engineering, Science and Technology (QuEST), Imperial College London, Prince Consort Rd, London, SW7 2AZ, United Kingdom}

\author{Yazeed K Alwehaibi}
\affiliation{Blackett Laboratory, Department of Physics, Imperial College London, Prince Consort Rd, London, SW7 2AZ, United Kingdom}
\affiliation{Centre for Quantum Engineering, Science and Technology (QuEST), Imperial College London, Prince Consort Rd, London, SW7 2AZ, United Kingdom}

\author{Oscar Scholin}
\affiliation{Cavendish Laboratory, University of Cambridge, Cambridge CB3 0FD, United Kingdom}

\author{Gerard J. Machado}
\affiliation{Blackett Laboratory, Department of Physics, Imperial College London, Prince Consort Rd, London, SW7 2AZ, United Kingdom}
\affiliation{Centre for Quantum Engineering, Science and Technology (QuEST), Imperial College London, Prince Consort Rd, London, SW7 2AZ, United Kingdom}

\author{Kuan-Cheng Chen}
\affiliation{Department of Electrical and Electronic Engineering, Imperial College London, Prince Consort Rd, London, SW7 2AZ, United Kingdom}
\affiliation{Centre for Quantum Engineering, Science and Technology (QuEST), Imperial College London, Prince Consort Rd, London, SW7 2AZ, United Kingdom}

\author{Aonan Zhang}
\affiliation{Blackett Laboratory, Department of Physics, Imperial College London, Prince Consort Rd, London, SW7 2AZ, United Kingdom}
\affiliation{Centre for Quantum Engineering, Science and Technology (QuEST), Imperial College London, Prince Consort Rd, London, SW7 2AZ, United Kingdom}
\affiliation{Clarendon Laboratory, University of Oxford, Parks Road, Oxford OX1 3PU, United Kingdom}

\author{Raj B Patel\footnote{raj.patel1@imperial.ac.uk}}
\affiliation{Blackett Laboratory, Department of Physics, Imperial College London, Prince Consort Rd, London, SW7 2AZ, United Kingdom}
\affiliation{Centre for Quantum Engineering, Science and Technology (QuEST), Imperial College London, Prince Consort Rd, London, SW7 2AZ, United Kingdom}

\author{Ying Dong\footnote{yingdong@cjlu.edu.cn}}
\affiliation{College of Metrology Measurement and Instrument, China Jiliang University, Hangzhou, 310018, China}

\author{Ian A. Walmsley}
\affiliation{Blackett Laboratory, Department of Physics, Imperial College London, Prince Consort Rd, London, SW7 2AZ, United Kingdom}
\affiliation{Centre for Quantum Engineering, Science and Technology (QuEST), Imperial College London, Prince Consort Rd, London, SW7 2AZ, United Kingdom}

\author{Vlatko Vedral}
\affiliation{Clarendon Laboratory, University of Oxford, Parks Road, Oxford OX1 3PU, United Kingdom}

\author{Ginestra Bianconi\footnote{ginestra.bianconi@gmail.com}}
\affiliation{Centre for Complex Systems, School of Mathematical Sciences, Queen Mary University of London, London, E1 4NS, United Kingdom}

\begin{abstract}
Understanding topological features in networks is crucial for unravelling complex phenomena across fields such as neuroscience, condensed matter, and high-energy physics. However, identifying higher-order topological structures---such as $k$-cliques, fundamental building blocks of complex networks---remains a significant challenge. Here we develop a universal programmable photonic quantum processor that enables the encoding of arbitrary complex-weight networks, providing a direct pathway to uncovering their topological structures. We demonstrate how this quantum approach can identify weighted $k$-cliques and estimate Betti numbers by leveraging the Gaussian boson sampling algorithm's ability to preferentially select high-weight, dense subgraphs. The unique capabilities of our programmable quantum processor allow us to observe topological phase transitions and identify clique percolation phenomena directly from the entropy of the sampling results. These findings showcase how photonic quantum computing can be applied to analyse the topological characteristics of real-world complex networks, opening new possibilities for quantum-enhanced data analysis.
\end{abstract}

\maketitle
\date{\today}

\noindent \textbf{Main}\\
A broad spectrum of scientific research employs complex networks to represent intricate interaction mechanisms \cite{Vespignani2018,Fortunato2022,Posfai2024,Bianconi2021}, helping illuminate diverse phenomena ranging from brain function \cite{Bullmore2009,Bassett2017}, protein interactions \cite{Wang2020,Meng2021,Qiu2023,Chen2024} to social dynamics \cite{Grewal2008,Lynn2020,Alvarez2021}. Extracting relevant information from the sheer complexity of networked data is a fundamental challenge of the field that requires the integration of approaches coming from different fields of science, including mathematics, physics and computer science.
In this context, topological data analysis (TDA) \cite{Baas2018,Sizemore2019,Carlsson2022,Harrington}, has emerged as a powerful framework, leveraging algebraic topology to uncover the underlying ``shape'' of data. Through a specific filtering process \cite{Baas2018,Carlsson2022} that systematically uncovers multi-scale topological structures, this approach identifies global characteristics in complex networks---such as Betti numbers and Euler characteristics. In this way, topological network analysis highlights stable, robust features, that other conventional methods overlook \cite{Carlsson2020}.
However, a major limitation of this methodology is its inherently high computational complexity (like computing Betti numbers is $\#P$-hard), leading to an exponential blow-up as the dataset size increases and rendering the analysis of massive networks computationally intractable in practice \cite{Basu1999,Basu2008,Lloyd2016,Schmidhuber2023}.

Quantum computing presents a novel platform to address this computational challenge with potential advantages over classical methods \cite{Harrigan2021,Berry2024}. Lloyd \emph{et al.} initially proposed a quantum machine learning algorithm for estimating Betti numbers \cite{Lloyd2016}, which was subsequently demonstrated on a photonic platform \cite{Huang2018}. Ongoing research continues to examine the advantages and limitations of quantum algorithms for computing different topological invariants in networks \cite{Akhalwaya2024,Crichigno2024,Gyurik2024,Gyurik2022Towards}, with recent results suggesting that while its quantum complexity may remain asymptotically hard, quantum algorithms still can provide a polynomial or even superpolynomial speedup over the best-known classical algorithms for certain types of input networks~\cite{Berry2024, Schmidhuber2023}.
In particular, the main computational bottleneck of TDA progress is the construction of the simplicial complex from the input network, which requires identifying all $k$ -cliques on the graph~\cite{Schmidhuber2023}. 

Gaussian boson sampling (GBS)~\cite{Lund2014,Hamilton2017,Zhong2020,Madsen2022,Yu2023} is a photonic quantum computing scheme that directly encodes an arbitrary complex-weighted graph into photon interference on a linear optical network. Photon-number measurements at the output effectively sample substructures of the graph with a probability distribution that favours higher subgraph-connectivity with heavier edge weights. GBS-based sampling algorithms have shown enhanced speed and performance over classical methods in dense-subgraph searching applications~\cite{Arrazola2022,Santi2022,Deng2023,Oh2024prxq}, precisely the structural characteristics one needs to capture for constructing the clique-complex. GBS therefore provides a useful tool for analysing datasets encoded as a complex network, that is also well suited for near-term noisy intermediate-scale quantum (NISQ) implementations. 

In this work, we leverage the statistical properties of GBS to develop a quantum approach for analysing topological features in complex networks. Implemented on a universal programmable photonic quantum computing system---featuring a tunable high-gain squeezed-light source and a fully reconfigurable interferometer---this platform is ideally suited for handling weighted network data. By encoding weighted networks into GBS, we identify $k$-cliques for computing Betti numbers and extract crucial weight information to guide a two-dimensional filtering process. This approach also enables us to compute the Euler characteristic from those cliques and track its evolution, revealing topological phase transitions \cite{Dorogovtsev2008,Santos2019}.

Leveraging the sensitivity of GBS to dense subgraphs, we further investigate clique percolation \cite{Derenyi2005} by analysing the entropy of the sampling outcomes, allowing us to probe critical phenomena characterized by abrupt changes in network connectivity. These results demonstrate how quantum techniques can uncover intricate topological properties that are difficult to capture classically. Given that many real-world problems can be represented as graph models, our photonic quantum approach provides a valuable framework for studying large-scale network structures, such as the structural complexity of brain networks \cite{Santos2019}. By effectively extracting higher-order topological structures from complex weighted networks, our work highlights the potential of quantum-enhanced methodologies in network science and beyond.

\begin{figure*}[hbt]
\centering
\includegraphics[width=0.98\textwidth]{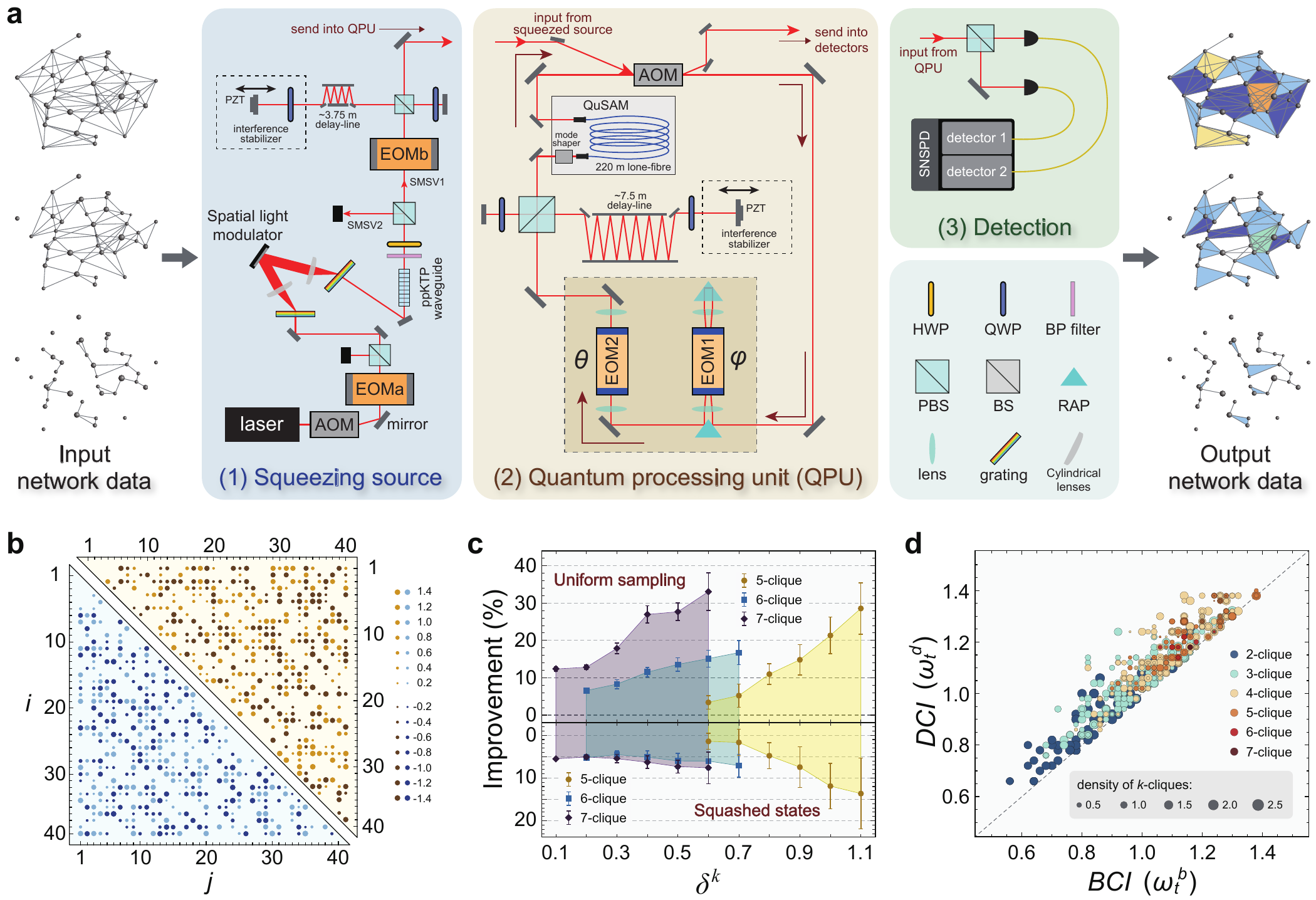}
\caption{\textbf{Photonic quantum processor for analysing networks with complex-weighted edges.}
\textbf{a,} The network data is fed into \emph{Babbage}, a universally programmable GBS machine employing temporal-polarization-
hybrid (TPH) encoding. The experimental output aids in identifying all $k$-cliques within the network. \emph{Babbage}, featuring expanded mode capacity, comprises three modules: (1) a squeezing source, (2) a quantum processing unit, and (3) a detection system. Details of the network encoding method and the experimental setup are provided in the Methods and Supplementary Information Section II. Abbreviations: periodically poled KTP (ppKTP), electro-optic modulator (EOM), acousto-optic modulator (AOM), half-wave plate (HWP), quarter-wave plate (QWP), (non-polarizing) beam splitter (BS), polarization beam splitter (PBS), right-angle prism mirrors (RAP), piezoelectric mirror (PZT), quantum sequential access memory (QuSAM) and superconducting nanowire single-photon detector (SNSPD).
\textbf{b,} The adjacency matrix of a dual-layer random network with complex weights. Dots in the yellow-shaded upper triangle represent the real part of the weighted edges, while those in the blue-shaded lower triangle represent the imaginary part. Lighter and deeper shades of yellow and blue correspond to positive and negative values, respectively.
\textbf{c,} Improvement in the success rate of finding $k$-cliques in the dual-layer network (Fig. 1b) using GBS, compared to uniform sampling and squashed-state sampling as benchmarks, for $5$-, $6$-, and $7$-cliques. Details can be found in the Methods.
\textbf{d,} By tracking the birth correlation intensity ($BCI$, $\omega_{t}^{b}$) and death correlation intensity ($DCI$, $\omega_{t}^{d}$) of $k$-cliques at different connectivity thresholds, this diagram illustrates the evolution of the network's topological structure. Sizes of circles highlight the density of $k$-cliques, while the distance away from the diagonal signifies their persistent features.
}
\label{Fig1}
\end{figure*}

\vspace{6pt}
\noindent \textbf{Analysing complex-number weighted networks with a photonic quantum computer.}\\
We developed a large-scale photonic quantum processor for analysing the topological features of networks. Using a resource-efficient, scalable temporal-polarization-hybrid (TPH) encoding scheme, this hardware enables us to embed arbitrary complex-weighted network data into a fully connected reconfigurable linear optical circuit equipped with adjustable squeezing elements. This design, realized in our device \emph{Babbage}, provides universal programmability and scalability---key requirements for translating theoretical quantum algorithms into practical applications.

\emph{Babbage} comprises three main modules (Fig. 1a): (1) Tunable squeezed source generator. An electro-optic modulator (EOMb) and a delay line equipped with an interference stabilizer \cite{Yu2020} ensure stable interference between adjacent pulses. (2) Quantum processing unit (QPU), fed with squeezed states, operates in both the temporal and polarization degrees of freedom \cite{Andersen2015,Takeda2017,Asavanant2019,Larsen2019,Enomoto2021,Yonezu2023,Yu2024}. This architecture allows arbitrary unitary transformations to be programmed through EOM1/2, implementing SU(2) operations \cite{Clements2016} on each time-bin mode. (3) After realizing unitary operations through sufficient loops in QPU, the pulses reach the detection module, where superconducting nanowire single-photon detectors (SNSPDs) record sampling results. Detailed parameters, including loss and fidelity, are provided in the Methods and Supplementary Information section II.

To validate \emph{Babbage}, we tested it on a random dual-layer network which contains both real and imaginary edge weights, respectively, shown in Fig. 1b. The edge between nodes $i$ and $j$ is represented by a dot whose size indicates its weight ($\omega_{ij}$). By incorporating these complex-valued edge weights---which represents correlations of varying intensities or phases \cite{Bottcher2024}, as seen in transmission-line networks \cite{Paul2007,Strub2020} and quantum networks \cite{Nokkala2024}---we capture more information beyond what non-negative, real-weighted models can achieve \cite{Bottcher2024}. 
This richer representation not only deepens our understanding of network structure and dynamics, but also provides a broader range of applications in quantum information, machine learning \cite{Kobayashi2016,Zhang2021,Spall2022}, and potentially in phase synchronization in complex systems \cite{Dai2021}. Crucially, GBS naturally biases its sampling toward higher-weight cliques, making it a useful tool for identifying these key substructures of complex network properties.

By comparing the performance of a clique-search heuristic algorithm (see Supplementary Information Section III and Ref. \cite{Banchi2020}) seeded with equal amount of GBS experimental data, classically uniform-sampled data \cite{Oh2024prxq}, and data sampled from squashed-state inputs \cite{Cifuentes2023}, we demonstrate that \emph{Babbage} outperforms these classical adversaries. The GBS enhancement is defined as the ratio of $k$-cliques identified by GBS to those found using classical methods. As illustrated in Fig. 1c for three different clique sizes, the enhancement becomes more pronounced at higher $k$-clique densities ($\delta^{k}$), which typically signify stronger interaction regions in real-world scenarios \cite{Bedru2020}. Here, $\delta^{k}$ is defined as $|\sum_{i,j=1}^{k} \omega_{ij}|/(k(k-1))$ \cite{Deng2023}, effectively treating the combined influence of complex edge weights as a quantum-like interference pattern \cite{Nokkala2024,Bottcher2024}. Further details are provided in Methods.

The identified $k$-cliques provide insight into the network's underlying high-order topological features \cite{Battiston2021}. As shown in Fig. 1d, by retaining only edges whose weight $|\omega_{ij}|$ is below a threshold $\omega_{t}$, we observe how specific weighted $k$-cliques emerge, persist, and eventually vanish. This time-resolved perspective offers a direct means of tracking how a network's topology evolves, enabling dynamic, real-time insights into complex systems, a practical example of which is presented in Supplementary Information Section II.

\begin{figure}[hbt]
\centering
\includegraphics[width=0.48\textwidth]{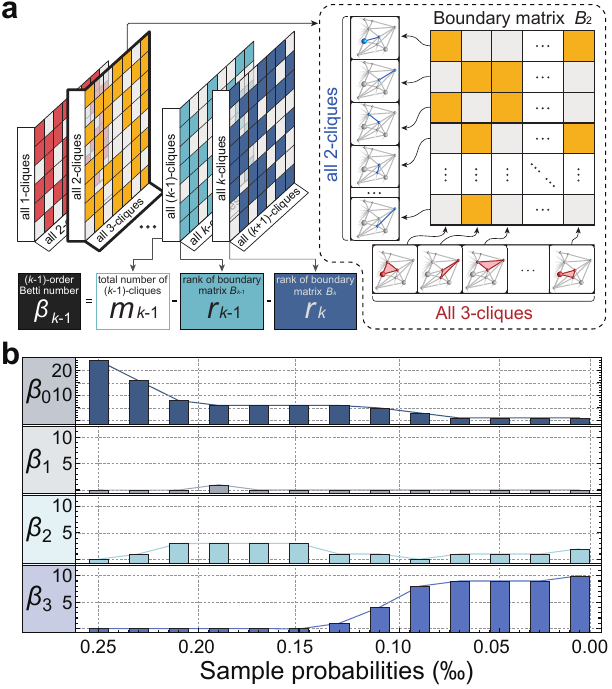}
\caption{\textbf{Capabilities of GBS in efficiently computing Betti numbers.} \textbf{a,} A flowchart illustrating the use of GBS to compute the Betti numbers of a network. The experimental results from GBS identify the $k$-cliques information. As shown in the dashed box on the right, $k$-cliques and ($k+1$)-cliques are used to construct the boundary matrix $B_{k}$. The element in $B_{k}$ is set to 1 if a $k$-clique is contained within a ($k+1$)-clique; otherwise, it is set to 0. By calculating the binary ranks of $B_{k}$ and counting the number of $k$-cliques, the Betti numbers can be determined.
\textbf{b,} Betti numbers computed by GBS using clique density as the filtering criterion. Since GBS tends to sample higher-weighted subgraphs, denser $k$-cliques are more likely to be identified from sampling events that exceed a chosen probability threshold, aided by post-processing. Without loss of generality, the filtering process here is then completed by reconstructing the network using those 5-cliques arising from sampling events above that threshold. Under this filtering procedure, changes in the Betti numbers ($\beta_{0-3}$) then provide insights into the network's underlying topological structure.
}
\label{Fig2}
\end{figure}

\vspace{6pt}
\noindent \textbf{Calculating Betti numbers using \emph{Babbage}.}\\
The identified $k$-cliques serve as a starting point for revealing a network's underlying topological features. For example, Betti numbers, $\beta_{k}$, which capture essential topological features such as connected components, loops, and other higher-dimensional ``holes'', are central to TDA and is computed from the $k$-clique information. For this purpose, we construct a boundary matrix $\boldsymbol{B}_k$ that encodes how lower-dimensional cliques (e.g., edges) assemble into higher-dimensional ones (e.g., triangles), as illustrated in Fig. 2a. By determining the binary ranks ($r_k$) of $\boldsymbol{B}_k$ and counting the number of $k$-cliques ($m_k$), we compute the Betti numbers via $\beta_k = m_k - r_k - r_{k+1}$ \cite{Shi2021}. 

We note that the complete search of all $k$-cliques of a graph remains a challenging task even for GBS, and in practice GBS needs to be combined with classical heuristic algorithms to reliably identify them (see Supplementary Information Section III). However, the output statistics of GBS highlight cliques with higher weights, which tend to be more relevant in real-world applications, while lower-weight cliques often warrant less attention in complex system analysis. As a result, when studying weighted networks, the output from GBS is particularly useful for pinpointing crucial, high-weight high-connectivity subregions.

To validate our technique, Fig. 2b presents the Betti numbers of various orders obtained from \emph{Babbage} after filtering cliques according to the absolute value of the clique density ($\delta^{k}$). High-density regimes commonly highlight regions of strong interactions or correlations in real-world networks, and this filtering process can naturally be guided by sampling probability. In Fig. 2b, we begin by reconstructing the network using the 5-cliques that exceed a certain density threshold, determined by sorting the probability distribution of GBS sampling results, and then compute all the necessary $k$-clique information for calculating $\beta_{k}$.

Our experimental results show that GBS can be used to extract Betti numbers, particularly in weighted networks, where a natural filtering method based on clique density is provided. These capabilities also extend to diverse applications---for instance, in protein engineering, where a nuanced understanding of structural dynamics can facilitate the identification of stable configurations and inform rational design \cite{Qiu2023,Wang2020}.

\begin{figure}[t]
\centering
\includegraphics[width=0.485\textwidth]{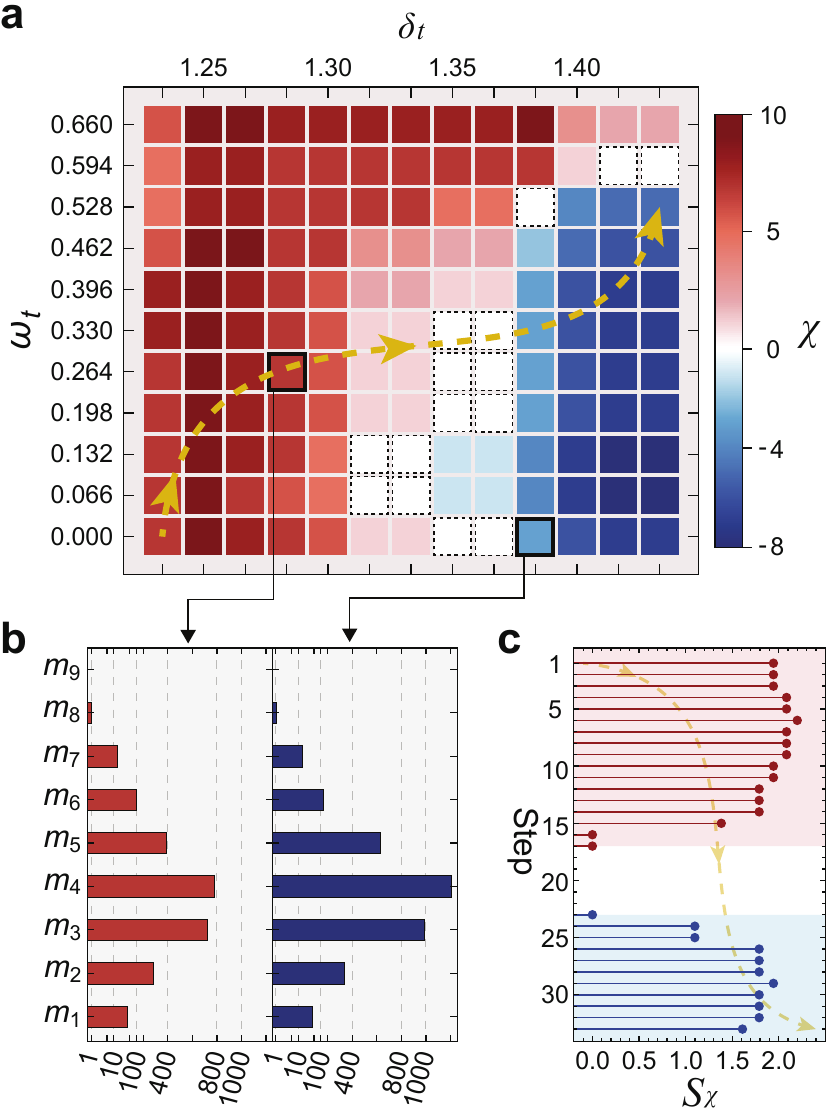}
\caption{\textbf{Topological phase transition observed with \emph{Babbage}.}
\textbf{a,} Variation of the Euler characteristic under a two-dimensional filtering process using both the correlation intensity threshold $\omega_{t}$ and the clique density threshold $\delta_{t}$, as introduced in Fig. 1d and Fig. 2b. The color scale represents changes in the Euler characteristic $\chi(\omega_{t},\delta_{t})$. Topological transition points occur at $\chi=0$, indicated by white squares with black dashed edges.
\textbf{b,} The clique information used to calculate $\chi$ at positions marked by black borders in \textbf{a}.
\textbf{c,} Detecting the TPT along a specified evolutionary path (yellow dashed line in \textbf{a}). Euler entropy ($S_{\chi}$) is measured at equal intervals during the evolution. A singularity where $S_{\chi}\to -\infty$ clearly signals the TPT region.
}
\label{Fig3}
\end{figure}

\begin{figure*}[hbt]
\centering
\includegraphics[width=1.0\textwidth]{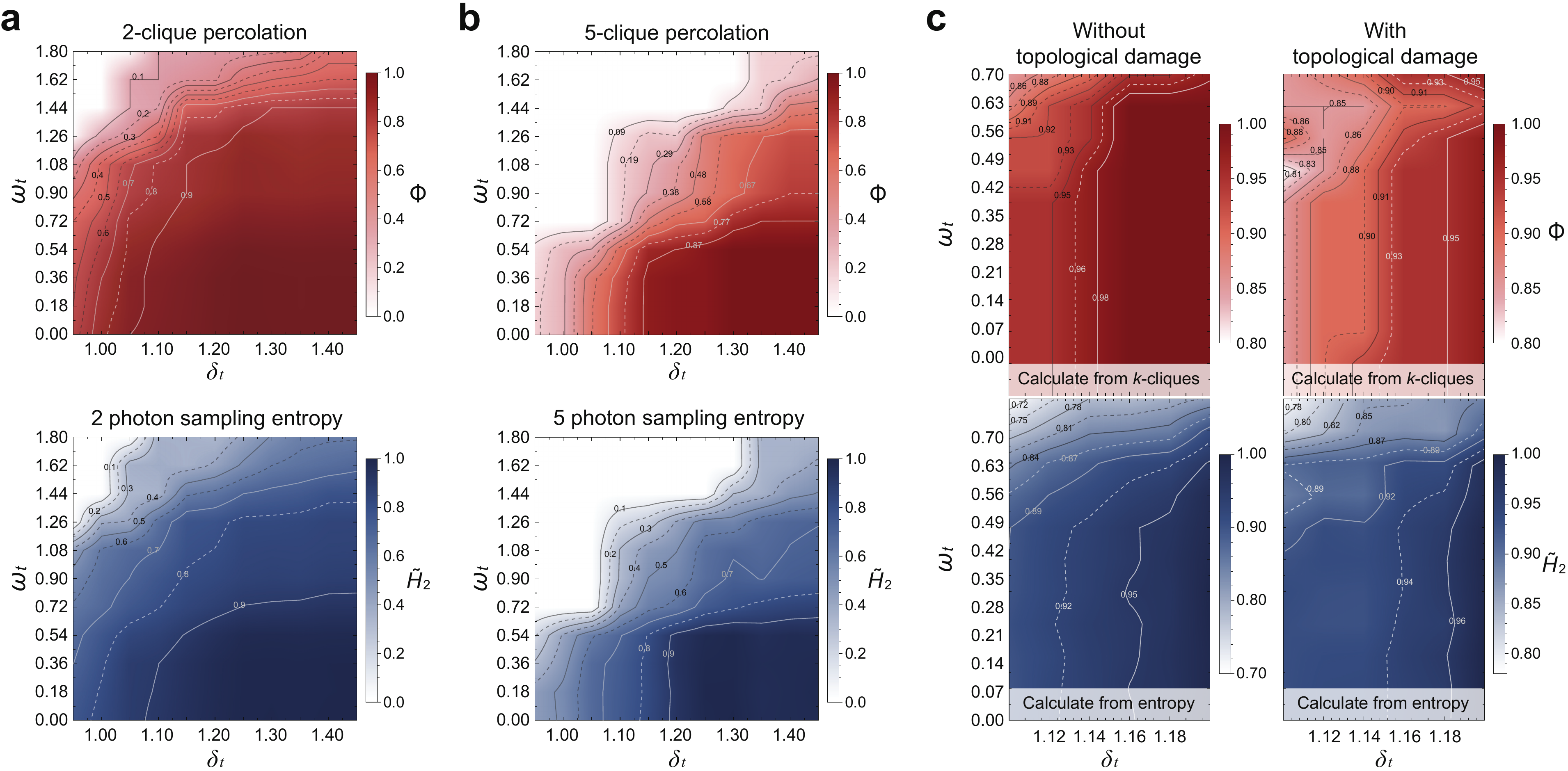}
\caption{\textbf{Detect percolation phenomenon using entropy of sampling pattern}. In \textbf{a,} and \textbf{b,} we measure the normalized R\'{e}nyi entropy (blue, lower panels) of the sampling photon probability distribution under a two-dimensional filtering process with two- and five-photon, respectively. Variations in R\'{e}nyi entropy correlate strongly with the order parameter $\Phi$ (red, upper panels), derived from clique-based analyses \cite{Derenyi2005}, indicating that the entropy measure can serve as a reliable percolation indicator. More detailed data are shown in Supplementary Information Section IV. \textbf{c,} The left column shows the undamaged scenario: $\Phi$ (red, upper panels) serves as the ground truth, while the normalized R\'{e}nyi entropy obtained from 4-photon sampling results (blue, lower panels) is displayed below. The topological damage shifts the R\'{e}nyi entropy, matching the changes indicated by the clique information (serving as the ideal reference), shown in the right column. This confirms the method's ability to detect topological damage.
}
\label{Fig4}
\end{figure*}

Despite challenges like photon loss \cite{Oh2024np}, ongoing advances in quantum hardware and algorithms can elevate GBS from a conceptual demonstration to a robust tool for exploring complex networks. Although a quantum computational advantage in estimating Betti numbers not exhibit, our main objective here is to show that GBS offers a novel quantum approach for searching simplicial complexes. Specifically, for complex weighted networks, there is currently no efficient classical method can simulate the outcomes of GBS experiments. In addition, gate-based quantum algorithms typically require substantial error-correction overhead, limiting their scalability. In contrast, GBS operates in a different way, which does not explicitly suffer from these constraints.

\vspace{6pt}
\noindent \textbf{Observing topological phase transition.}\\
A distinctive hallmark of complex networks is the abrupt shift in their global structure as certain parameters vary. TDA captures these phenomena by examining changes in the Euler characteristic ($\chi$), a key topological invariant that encodes the global connectivity and the presence of loops or voids \cite{Ghrist2016}. Using our device \emph{Babbage}, we performed a two-dimensional filtering process on the complex weight network data, identifying the number of $k$-cliques (Fig. 3b) and then computing $\chi=\sum_{k=0}(-1)^{k}m_{k}$ to pinpoint topological phase transition (TPT) points---crucial parameters at which $\chi$ vanishes, indicating fundamental shifts in the network's overall structure, as shown in Fig. 3a.

We further demonstrated that by moving along a specified path in this two-dimensional TDA process (the yellow dashed line in Fig. 3a), the system's TPT point could be experimentally determined. Leveraging Euler entropy, defined as $S_{\chi}=\ln(|\chi|)$, we successfully identified the onset of these transitions (Fig. 3c). Beyond providing a robust measure of structural change in network topology, this approach can serve as a clinically relevant topological biomarker. For example, it can distinguish between healthy individuals and glioma patients \cite{Santos2019}, underlining its potential impact in real-world medical diagnostics.

\vspace{6pt}
\noindent \textbf{Detecting clique percolation with sampling entropy.}\\
Having established a method to compute Betti numbers and pinpoint topological phase transitions, we turn next to another critical phenomenon---$k$-clique percolation \cite{Derenyi2005}---to further explore the connectivity of higher-order structures in complex networks \cite{Sun2023}. A $k$-clique percolation cluster is a maximal $k$-clique-connected subgraph composed of $k$-cliques that share $k-1$ nodes in series. To quantify the clique percolation, an order parameter $\Phi$ is defined as $\Phi = N^{*}/N$, where $N^{*}$ is the number of nodes in the percolation cluster and $N$ the total number of nodes in the network \cite{Derenyi2005}. More details are given in Methods and Supplementary Information Section V. 

Identifying the percolation threshold by enumerating every $k$-clique is conceptually straightforward (as shown in Fig. 4a,b, upper panels in red, serving as the ideal benchmark values here). However, this approach can be computationally challenging for classical algorithms \cite{Palla2005} and even for quantum computers \cite{Schmidhuber2023}. We explore how to streamline this process by leveraging the statistical properties of the GBS output distribution. Because clique percolation fundamentally captures the essence of network connectivity, analysing GBS outcomes with an entropy metric \cite{Nokkala2018} such as R\'{e}nyi entropy ($H_{\alpha} = (1/(1-\alpha)) \log\sum p_{i}^{\alpha}$), which highlights more significant events while dampening rarely occurring patterns and noise, provides a meaningful qualitative view of how connections distribute themselves and can potentially uncover key percolation features.

Higher $H_{\alpha}$ values indicate a denser, more uniform edge distribution and thus better connectivity, whereas lower entropy signifies the onset of fragmentation. In our tests with various $k$-cliques percolation procedure, we observed that the normalized R\'{e}nyi entropy $\tilde{H}_{2}$ for $k$-photon sampling patterns closely mirrors the percolation behavior, as shown in Fig. 4a and b, lower panels with blue colour. Crucially, this approach avoids the need for exhaustive enumeration of all $k$-cliques, as changes in $H_{2}$ remain in close agreement with the clique-based measure $\Phi$. Hence, it serves as a sensitive percolation indicator and bolsters the practical utility of our quantum method for unraveling complex networks. Further details are provided in Supplementary Information Section V.

Furthermore, we use this approach to examine topological damage \cite{Bianconi2018pre}, highlighting how localized disruptions propagate through higher-order structures---thereby revealing a network's resilience and fragility, which can be detected by \emph{Babbage}. As a practical demonstration, we deliberately removed edges within the 4-cliques containing node 1, inducing corresponding shifts in the R\'{e}nyi entropy (lower panels, Fig. 4e). These shifts closely tracked the ground-truth order parameter $\Phi$ computed from clique information (upper panels, Fig. 4e), confirming that this entropy-based measurement can reliably detect this kind of topological damage. Notably, this approach naturally extends to larger, more complex networks and can be applied dynamically, allowing percolation thresholds to be identified on-the-fly as network parameters evolve.

\vspace{6pt}
\noindent \textbf{Discussion and outlook.}\\
Our demonstration of topological network analysis on a universal programmable photonic quantum processor adapts readily available quantum technologies to analysing complex-weighted networks, a computationally challenging task but with important scientific and industrial applications. By exploiting GBS's inherent bias toward higher-weight $k$-cliques, our method naturally targets weighted substructures. In networks with uneven weight distributions or a few large, high-value cliques, this bias substantially narrows the search space by focusing on the most promising regions first. Consequently, GBS acts as a probabilistic guide that can uncover key cliques faster than uniform or other quantum-inspired classical methods, particularly in large-scale or heavily weighted networks. 
By leveraging these features, we are able to investigate topological phase transitions and clique percolation phenomena by tracking how $k$-cliques aggregate and percolate across varying thresholds.
Moreover, because of the natural selection of dominant substructures such as cliques and tracking how they percolate under varying thresholds, one may uncover the critical phenomenon that governs large-scale behaviour in brain activity \cite{Santoro2024}, robust conduction in materials \cite{Hasan2010}, or connectivity of porous medium \cite{Thakur2021}.

\emph{Babbage} features the largest number of modes with a universal programmable design, making it a powerful platform for extracting high-order topological invariants in real-world weighted network scenarios. Additionally, its versatility extends to applications such as cluster state generation and measurement-based quantum computing algorithms. Moving forward, our roadmap with a modular architecture is designed to support both multi-core and distributed implementations. Furthermore, it enables the integration of non-Gaussian operations \cite{Namekata2010, Konno2024}, making the platform well-suited for fault-tolerant photonic computing \cite{Putterman2025}. This, in turn, allows for the exploration of high-order interactions and nonlinear dynamics in higher-order networks \cite{Millan2025}. By leveraging these advancements, we anticipate steady progress toward scaling to hundreds of modes, enhancing the potential for quantum advantage in network-related applications.

\vspace*{1\baselineskip}

\begin{acknowledgments}
{\bf Acknowledgments}

We thank Nicol\'{a}s Quesada and Changhun Oh for the helpful discussions and feedback on the manuscript. This work was supported by UK Research and Innovation Guarantee Postdoctoral Fellowship (project: EP/Y029631/1), Engineering and Physical Sciences Research Council (EPSRC) UK Quantum Technologies Program's hubs for Quantum Computing \& Simulation (project: EP/T001062/1) and Quantum Computing via Integrated and Interconnected Implementations (project: EP/Z53318X/1), Schmidt Sciences, LLC., UK Research and Innovation Future Leaders Fellowship (project: MR/W011794/1), EU Horizon 2020 Marie Sklodowska-Curie Innovation Training Network (project: 956071, `AppQInfo') and National Research Council of Canada (project QSP 062-2).

\end{acknowledgments}

{\bf Contributions}

S.Y. conceived the experiment, and J.S. conceived of the theoretical model. S.Y. built the experimental setup with the help of R.B.P. and Y.D. S.Y., J.S. and Z.L. completed data analysis with the help of E.M., Y.K.A., O.S., G.J.M., K.-C.C. and A.Z. S.Y., S.J. and Z.L. prepared the manuscript with the help of E.M., Y.K.A., O.S., G.J.M., K.-C.C. R.B.P., Y.D., I.A.W., V.V, and G.B. All authors discussed the results and reviewed the manuscript.

{\bf Competing interests}

The authors declare no competing interests.

\vspace*{1\baselineskip}

{\bf Methods}

\emph{Details about TDA}---
The topological structure of a network profoundly influences its functionality, efficiency, and the dynamics of information flow. Simple connectivity patterns, such as star, ring, or mesh topologies, can determine how quickly and reliably information propagates. Yet, as networks grow in complexity---spanning domains from biomedical research to climatology---traditional analytical methods often fail to capture their intricate, high-dimensional topological features.

TDA is a framework that employs concepts from algebraic topology to understand the ``shape'' of complex datasets. Rather than focusing solely on pairwise interactions or statistical correlations, TDA seeks to identify structural patterns and multi-scale features hidden within the data. This approach goes beyond conventional methods by examining how data points connect to form loops, voids, and higher-dimensional cavities. By quantifying such features, TDA provides insights into underlying organizational principles that govern system behavior.

A cornerstone concept in TDA is the Betti number---a quantifier of the number of topological ``holes'' in various dimensions within a dataset. These holes offer insight into connectivity, redundancy, and modular organization. For example, $\beta_{0}$ reflects the number of connected components, shedding light on the network's overall fragmentation or cohesion. $\beta_{1}$ counts one-dimensional loops, pointing to potential communication pathways or redundant links. Higher-order Betti numbers, such as $\beta_{2}$ and beyond, capture even more intricate multi-dimensional features---like voids and higher-dimensional cavities---revealing complex structural patterns that influence how the network functions and evolves.

By analyzing topological invariants like Betti numbers, TDA enables a more profound understanding of network behavior. However, computing Betti numbers is a $\#$P-hard problem \cite{Schmidhuber2023}, highlighting fundamental challenges in scaling TDA to large, complex networks. These computational difficulties underscore the pressing need for innovative solutions capable of handling the intricate structures that increasingly define modern data landscapes.

\emph{Details about the TPH-encoding photonic quantum processor}---
A schematic diagram of the experimental setup for the universal programmable, and highly scalable TPH-encoding photonic quantum processor, \emph{Babbage}, is shown in Fig. 1a in the main text. The TPH encoding method first shown in the squeezing source module (1), depicted in the blue box. After generating the single-mode squeezed vacuum (SMSV), an electro-optical modulator b (EOMb) with digital amplifier (orange box with grey edge shown in Fig. 1a) is used to adjust the polarization of the SMSV. Following a 7.5 m delay line equipped with an interference stabilizer, two adjacent SMSVs (i.e., two adjacent time-bin modes) fall into the same time slot. The transmission rate of this module is approximately $70\%$.

In the QPU module, all time-bins are directly coupled into the loop through an AOM. By leveraging the overlap of two adjacent modes in the same time slot in module (1), we reduced the use of two EOMs as optical switches \cite{Yu2023}, thereby enhancing scalability and minimizing loss errors. Then, a ``core'' as shown in the dashed-dark-yellow box, enabling SU(2) control on two adjacent time bin modes. The core consisted a EOM pair (with analogue amplifier, and shown as light orange box with blue edge in Fig. 1a) for adjusting phase ($\varphi$, EOM1) and amplitude ($\theta$, EOM2) to achieve SU(2) operation.

The core structure introduced here allows us to better understand the modular composition of the QPU. It also inspires the possibility of placing more cores within the loop, enabling us to perform more SU(2) operations within a single round. This multi-core construction method can increase the depth within the same number of rounds, significantly enhancing the scalability of the device.

A delay-line device then interlaces pulses of different polarizations between adjacent time-bins, implementing Clements' architecture. This TPH encoding and compact design are particularly well-suited for scenarios such as GBS, where operations are required only between two adjacent time-bins. It is worth noting that an interference stabilizer \cite{Yu2020} is inserted in the delay line device both in modules (1) and (2) for maintaining the stability of photonic quantum processor for long-time data collection. The QuSAM can achieve an overall delay of approximately 1100 ns for each photon. A mode-shaper consisting of a 4f-system is built before the light is coupled into the 220 m long fibre. The operating frequency of \emph{Babbage} is 20kHz, and the total transmission efficiency for one loop is around $86.85\%$. More details can be found in Supplementary Information Section II.

After completing the same number of cycles as the number of modes, this QPU module ensures full connectivity for each time-bin mode, enabling the encoding of arbitrary networks. Assuming $A$ is the adjacency matrix of the network, we can decompose it as $A'=U\oplus_{i}^{N}\tanh{r_{i}}U^{T}$, where $A'$ is rescaled as $A'=cA+dI$ to ensure that all eigenvalues $\tanh{r_{i}}$ lie between 0 and 1 \cite{Banchi2020}. Here, $U$ represents the unitary matrix implemented in the QPU, and $r_{i}$ corresponds to the squeezing parameters achieved in the squeezing source. Finally, all time-bins are sent to the detector module to obtain sampling results. 

This architecture features a modular design in which the free-space delay line, the EOM-pair-based core, and the long-fiber delay components in the QPU are relatively independent. This modularity enables the addition of different modules (such as the ``core'') within the loop more flexible and provides strong scalability for the entire setup.

\emph{Complex weight dual-layer network}---
Complex weights in networks incorporate both magnitude and phase information, enabling encoding for information and precise characterization of dynamic behaviors, while providing a powerful mathematical framework to capture coherence and interference phenomena in the topology, making them essential for exploring high-dimensional features and quantum effects in networks. 

Fig. 1b provides a detailed representation of the adjacency matrix corresponding to the network we used in this work. 
The complex weight of each edge connecting nodes $i$ and $j$ is denoted as $\omega_{ij}=\alpha + i \beta$.
Then, the network is divided into two layers based on the real and imaginary parts of these weights. The upper right portion of the adjacency matrix, represented by yellow circles, corresponds to the real-part layer, where the size of each circle reflects the magnitude of the real part ($\alpha$). The lower left portion, represented by blue circles, corresponds to the imaginary-part layer, with their size indicating the magnitude of the imaginary part ($\beta$). As the network is undirected, the adjacency matrix is symmetric, and only half of it is displayed here.

This complex weight can be regarded as the correlation between nodes $i$ and $j$. Since the edge weights are complex numbers, this correlation captures both the magnitude and phase relationships of the node weights, offering a comprehensive measure of their interaction.

Additionally, it is worth noting that in GBS experiments, even though the error for large-weight subgraphs identified by GBS becomes larger within a small-scale weight range (see Fig. 1c in \cite{Deng2023}), there remains a positive correlation between the sampling outcomes and the discovery of higher-density subgraphs when the network possesses complex-valued edge weights \cite{Deng2023}. By taking weights into account in a suitable post-processing procedure \cite{Banchi2020}, one can still leverage GBS results to uncover a greater number of higher-density cliques compare with classical data input, as shown in Fig.1 c.

\emph{Improvement in the success rate of GBS compared to classical sampling}---
The advantage of using GBS for graph problem analysis comes from the fact that using GBS's experimental results to assist in the search of higher weighted (or higher density) cliques is more efficient than classical methods. Here, we analyze and explain the efficiency compare with classical methods.

Recent studies have proposed using classical methods to benchmark the experimental results of GBS. Specifically, Ref. \cite{Oh2024prxq} suggests that for non-negative graphs, the success rate of finding cliques using quantum-inspired classical methods can closely approximate the success rate achieved by GBS experiments. This occurs primarily because encoding a non-negative graph into a GBS device does not involve photon phase interference in the linear interferometer, thereby losing some quantum characteristics and enabling classical methods to effectively simulate the experimental results.

In our work, which considers networks with complex weights, thus comparing the GBS results with classical uniform sampling remains relevant. Incorporating both real and imaginary layers into a dual-layer network allows for a more nuanced representation of complex relationships. Real-valued edges can capture conventional interaction strengths or correlations, while imaginary-valued edges introduce phase-like information that can reflect directional influences, temporal delays, or oscillatory behaviours.

Success rate $p$ of finding the $k$-clique after post-processing (i.e., greedy shrinking and local search \cite{Banchi2020}) is obtained by calculating the proportion of the number of all correct $k$-cliques found in the total input sampling quantity, namely, $p_{k-\text{cliques}}=N_{k-\text{cliques}}/N_{\text{total samples}}$. Here, the samples are generated from GBS experiments, and compared with the uniform classical approach and squashed state input case. To demonstrate the efficiency advantage of GBS, we generated approximately 3000 samples in both the GBS experiments and numerical simulations on a classical computer. These samples were then processed through a post-processing program to determine the number of $k$-cliques found, allowing us to calculate $p_{\text{GBS}}$, $p_{\text{uniform}}$ and $p_{\text{squashed}}$. The improvement then is obtained by $p_{\text{GBS}} / p_{\text{classical}}$ and $p_{\text{GBS}} / p_{\text{squashed}}$, shown in Fig. 1c.

\emph{Boundary matrix and Betti number}---
For the $k$-cliques information we identified by \emph{Babbage}, let 
$\{\sigma_1, \sigma_2, \dots, \sigma_m\}$
be the set of all $k$-cliques, and 
$\{\tau_1, \tau_2, \dots, \tau_n\}$
be the set of all $(k-1)$-cliques. The boundary matrix $B_{k}$ is an $n \times m$ matrix, where each entry $B_k(i, j)$ is defined as:
\[
\boldsymbol{B}_k(i, j) =
\begin{cases} 
1 & \text{if } \tau_i \text{ is included in } \sigma_j, \\
0 & \text{otherwise}.
\end{cases}
\]

In Fig. 2b, without loss of generality, we use the absolute value of the 5-clique density, $|\delta_{t}^{k=5}|$, as the threshold in the filtering process. Specifically, we first plot the 5-photon GBS sampling pattern, where higher-probability regions typically indicate denser cliques that are more likely to be found. We then set a sampling probability threshold and, for all sampling results whose probabilities exceed that threshold, identify the corresponding 5-cliques and reconstruct the graph accordingly.

\emph{Topological phase transition}---
In Fig. 3a, we conducted a two-dimensional TDA analysis by incorporating both correlation and density as filtering parameters \cite{Santos2019}. This combined approach provides a more comprehensive understanding of topological phase transitions. Correlation measures the local strength of interactions between node pairs, illustrating how strongly certain edges bind the network at a microscopic level. Density, on the other hand, characterizes the global robustness and cohesiveness of subgraphs, revealing how clusters of nodes persist or dissolve as the system evolves. Here, we specifically use the density of 5-cliques as the filtering value. By examining topology through these two lenses simultaneously, researchers can determine whether critical phenomena arise from local connectivity shifts or from the formation and breakdown of highly cohesive substructures. This dual-parameter strategy thus connects micro-level interaction strengths to macro-level structural stability, offering a richer, multi-scale perspective on how networks undergo topological changes and enabling more nuanced insights into complex systems across diverse applications.

Moreover, GBS sampling naturally favours regions of higher clique density (especially in the positive weighted network)---precisely those areas of the parameter space that hold the greatest interest in topological data analysis. Since our dual-parameter filtering framework already highlights these dense, topologically significant substructures, the synergy between GBS sampling and TDA becomes particularly pronounced. By directing attention toward the most robust and cohesively connected regions of the network, GBS enhances our ability to identify meaningful patterns, track critical transitions, and gain deeper insights into the underlying dynamics and organization of complex systems.

\emph{Details about using R\'{e}nyi entropy to detect clique percolation}---
Here, we first briefly review the definition about the order parameter calculated with relative size of the largest $k$-clique percolation cluster \cite{Derenyi2005}.
In this framework, two $k$-cliques are considered adjacent if they differ by exactly one node. By connecting such adjacent $k$-cliques into chains, we identify $k$-clique percolation clusters---maximal sets of $k$-cliques linked together through shared nodes \cite{Derenyi2005} (see Supplementary Information Section V for details). To quantify the relative size of the largest $k$-clique percolation cluster, we define $\Phi = N^{*}/N$, where $N^{*}$ is the number of nodes in that cluster and $N$ is the total number of nodes in the network \cite{Derenyi2005}.

R\'{e}nyi entropy is an effective metric for analysing the distribution of GBS sampling results, offering insights into the diversity and concentration of sampled patterns. In R\'{e}nyi entropy, the parameter $\alpha$ determines how the measure weights events of different probabilities. Higher $\alpha$ emphasizes high-probability outcomes, while lower $\alpha$ highlights low-probability events. As $\alpha$ increases, the measure becomes more sensitive to dominant contributions in the sampling distribution. This attribute renders R\'{e}nyi entropy a powerful tool for understanding the characteristics of GBS outputs.

In Fig. 4, for the 2-photon results shown in Fig. 4a, we collected approximately 16,000 samples in total to establish the sampling distribution used for calculating R\'{e}nyi entropy. Due to losses in the setup, 5-photon events occurred less frequently. Consequently, for Fig. 4b, we obtained approximately 3000 samples. To clearly illustrate the differences in percolation thresholds, we used the density of 4-cliques here in the filtering process. Besides, we want to notice that the GBS-based sampling strategy naturally emphasizes high density cliques---ensuring the resulting distribution better reflects practical flow or propagation phenomena.

\emph{Advantage and complexity analysis}---
While identifying all $k$-cliques is $\#$P-hard in the worst case---making a full, exhaustive search inevitably exponential---GBS offers practical benefits under realistic conditions. In contrast to previous classical algorithms that mainly address clique detection in unweighted networks, our focus on complex-weighted networks remains underexplored in classical approaches. Because GBS sampling inherently biases detection toward high-weight or densely connected substructures, it can act as a probabilistic guide for locating crucial cliques more quickly than uniform or brute-force classical methods. In networks with uneven weight distributions or where certain large, high-weight cliques dominate, this property substantially reduces the search space by focusing on the most promising regions first. Although GBS cannot eliminate the fundamental complexity of enumerating every clique, it could serve as a promising candidate to streamline the process in typical or specialized scenarios. We discussed this in Supplementary Information Section III.
Moreover, hybrid approaches, where GBS pinpoints likely candidate cliques and classical algorithms verify and refine those findings, further enhance efficiency. As a result, GBS presents a practical route to speed up clique detection, especially in large-scale and complex weighted networks. This approach not only pushes the boundaries of clique detection but also encourages the development of new classical methods and associated complexity studies.

Meanwhile, although there has been active research on simulating GBS with classical hardware \cite{Oh2024prxq,Oh2024np} and on applying quantum algorithms to TDA tasks \cite{Lloyd2016}, our approach provides distinct benefits over existing approaches in terms of scalability, generality, and utility for large or complex datasets. Below, we elaborate on these benefits.
Notably, quantum-inspired approximation methods handle non-negative weighted graphs efficiently \cite{Oh2024prxq}, but this assumption can become restrictive when networks include complex-weighted edges. In contrast, our method directly encodes arbitrary complex-weighted adjacency matrices, addressing a broader class of problems. Even if one were to force non-negative weights for simpler tasks like clique identification, quantum-inspired techniques indicate that GBS continues to excel at identifying high-probability events (see Fig. 4 in \cite{Oh2024prxq}), which are most relevant for uncovering significant $k$-cliques. This inherent capability ensures that GBS remains advantageous, particularly in capturing statistically dominant substructures that shape a network's core topology.

Moreover, recent developments in tensor-network-based classical simulations have established strong baselines for approximating GBS outputs under lossy conditions \cite{Oh2024np}. While these simulations pose a challenge to GBS at small scales, Ref. \cite{Oh2024np} also indicates that classical approximations grow sharply in complexity with increasing modes. By contrast, our GBS setup features built-in scalability, allowing it to operate at mode counts that surpass the limits of these classical simulations, thereby facilitating genuine quantum computational gains. As our system extends to hundreds, or even thousands, of modes, classical simulation becomes intractable, paving the way for practical quantum advantages under realistic experimental conditions.

From a complexity-theoretic perspective, TDA algorithms---especially those aimed at extracting topological invariants like Betti numbers---are unlikely to be efficiently simulated by classical means, preserving potential quantum advantages~\cite{Gyurik2022Towards}. Certain parameter regimes may deliver  polynomial speedups \cite{Schmidhuber2023}, while others promise superpolynomial or even exponential improvements \cite{Berry2024}. Although our work does not focus on demonstrating a practical quantum advantage---partly because our hardware remains relatively small, and because enumerating all $k$-cliques or precisely computing Betti numbers are $\#P$-hard problems that prevent exponential speedups even on quantum computers---our approach still provides an promising framework for simultaneously obtaining both the identities and weights of $k$-cliques. This capability benefits the filtering process and $k$-clique searches in TDA, an avenue not previously explored.

\end{document}